# Vowel recognition with four coupled spin-torque nano-oscillators

Miguel Romera[1†], Philippe Talatchian[1†], Sumito Tsunegi[2], Flavio Abreu Araujo[1,‡], Vincent Cros[1], Paolo Bortolotti[1], Juan Trastoy[1], Kay Yakushiji[2], Akio Fukushima[2], Hitoshi Kubota[2], Shinji Yuasa[2], Maxence Ernoult[1,3], Damir Vodenicarevic[3], Tifenn Hirtzlin[3], Nicolas Locatelli[3], Damien Querlioz[3*], Julie Grollier[1*]

[1] - Unité Mixte de Physique, CNRS, Thales, Univ. Paris-Sud, Université Paris-Saclay, 91767 Palaiseau, France

[2] - National Institute of Advanced Industrial Science and Technology (AIST), Spintronics Research Center, Tsukuba, Ibaraki 305-8568, Japan

[3] - Centre de Nanosciences et de Nanotechnologies, CNRS, Univ. Paris-Sud, Université Paris-Saclay, 91405 Orsay France

[‡] - Now at: Institute of Condensed Matter and Nanosciences, UCLouvain, Place Croix du Sud 1, 1348 Louvain-la-Neuve, Belgium

[†] These two authors have equally contributed to the work

[*] julie.grollier@cnrs-thales.fr, damien.querlioz@u-psud.fr

In recent years, artificial neural networks have become the flagship algorithm of artificial intelligence[1]. In these systems, neuron activation functions are static and computing is achieved through standard arithmetic operations. By contrast, a prominent branch of neuroinspired computing embraces the dynamical nature of the brain and proposes to endow each component of a neural network with dynamical functionality, such as oscillations, and to rely on emergent physical phenomena, such as synchronization[2–7], for computing complex problems with small size networks[7–11]. This approach is especially interesting for hardware implementations, as emerging nanoelectronic devices can provide highly compact and energy-efficient non-linear auto-oscillators that mimic the periodic spiking activity of biological neurons[12–16]. The dynamical couplings between oscillators can then be used to mediate the synaptic communication between neurons. However, one major challenge towards implementing these models with nano-devices is to achieve learning, which requires finely controlling and tuning their coupled oscillations[17]. The dynamical features of nanodevices can indeed be difficult to control, and prone to noise and variability[18]. In this work, we show that the outstanding tunability of spintronic nano-oscillators, i.e. the possibility to widely and accurately control their frequency through electrical current and magnetic field, can solve this challenge. We successfully train a hardware network of four spin-torque nano-oscillators to recognize spoken vowels by tuning their frequencies according to an automatic real-time learning rule. We show that the high experimental recognition rates stem from the outstanding ability of these oscillators to synchronize. Our results demonstrate that non-trivial pattern classification tasks can be achieved with small hardware neural networks by endowing them with non-linear dynamical features: here, oscillations and synchronization. This demonstration of real-time learning with an array of four spin-torque nano-oscillators is a milestone for spintronics-based neuromorphic computing.

Spin-torque nano-oscillators are natural candidates for building hardware neural networks made of coupled nanoscale oscillators[8–10,13,15,18,19]. These nanoscale magnetic tunnel junctions emit microwave voltages when they are driven by dc current injection in a regime of sustained magnetization precession through the effect of spin torque. In addition, they have exceptional capacities to synchronize their rhythms to periodic electric and magnetic input signals and to other spin-torque nano-oscillators[20–24]. This property originates from the high tunability of their frequency, in other words, the large frequency changes induced by applied dc currents and magnetic fields. It has been recently demonstrated that single spin-torque nano-oscillators can achieve impressive cognitive computations[25]. However, it has not been shown experimentally that a coupled network of spin-torque nano-oscillators can learn to perform computational tasks through synchronization. Here, we use the ability of spin-torque nano-oscillators to modify their frequency in response to injected dc currents to train in real-time a network of coupled oscillators to categorize different input patterns into different synchronization configurations[2,17,18].

We transpose to hardware the neural network illustrated in Fig. 1a[17] with the set-up illustrated in Fig. 1b. The four neurons in Fig. 1a are experimentally implemented with four spin-torque nano-oscillators (Fig. 1b), in our case circular magnetic tunnel junctions with 375 nm diameter and an FeB free layer with a vortex as ground state (see Methods) [26]. The double arrow connections between neurons (blue in Fig. 1a) indicate that the output of neuron *i* influences the behavior of neuron *j*, and vice versa. We implement these symmetric neural interconnections by connecting electrically the four oscillators using millimeter-long wires as schematized in Fig. 1b: in this configuration, the microwave current generated by each oscillator propagates in the electrical microwave loop and in turn influences the dynamics, and in particular the frequency, of the other oscillators through the microwave spin-torques it creates[24]. The sum of all microwave emissions is detected by a spectrum analyzer. Importantly, we can control the frequency of each oscillator by adjusting the dc current flowing through each (see Methods and Extended Data Fig. 1). Here, for computing, we choose dc currents leading to close but not identical frequencies. The light blue curve in Fig. 1c shows a four-peak spectrum typical of this regime of moderate coupling where the dynamics of the oscillators are correlated but do not lead to mutual synchronization.

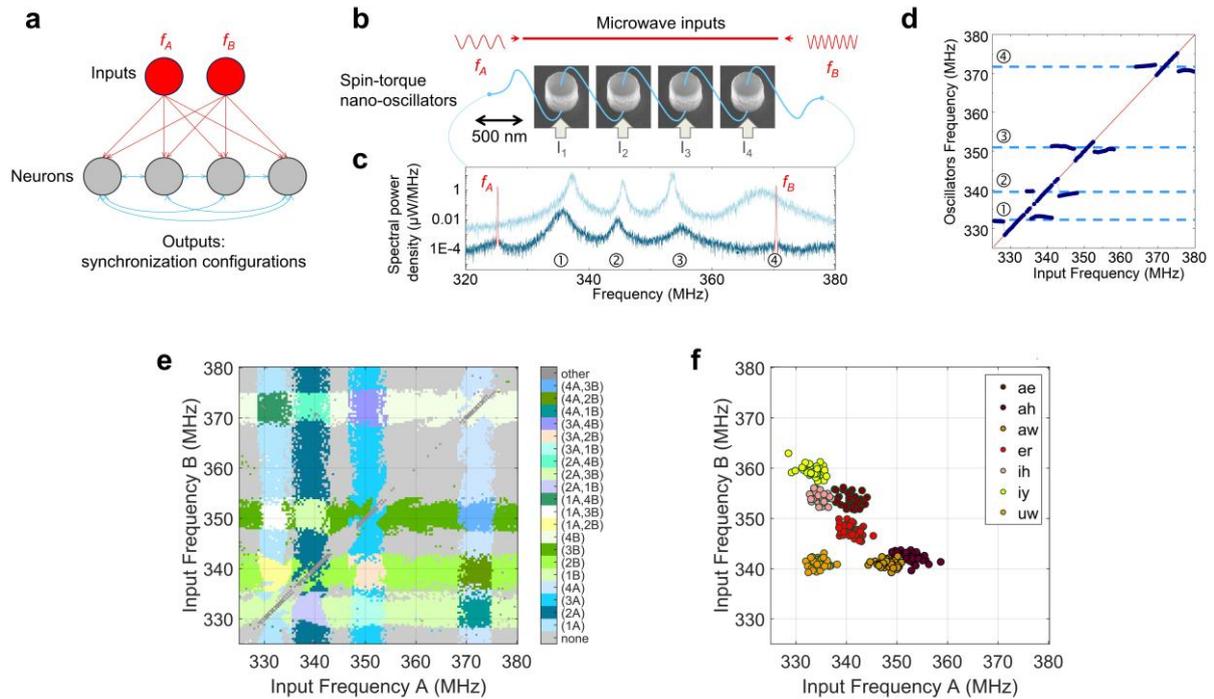

**Fig. 1. Approach for pattern classification with coupled spin-torque nano-oscillators. (a)** Schematic of the emulated neural network. **(b)** Schematic of the experimental set-up with four spin torque nano-oscillators electrically connected in series and coupled through their own emitted microwave currents. Two microwave signals encoding information in their frequencies $f_A$ and $f_B$ are applied as inputs to the system through a strip line, which translates into two microwave fields. The total microwave output of the oscillator network is recorded with a spectrum analyzer. **(c)** Microwave output emitted by the network of four oscillators without (light blue) and with (dark blue) the two microwave signals applied to the system. The two curves have been shifted vertically for clarity. The four peaks in the light blue curve correspond to the emissions of the four oscillators. The two red narrow peaks in the dark blue curve correspond to the external microwave signals with frequencies $f_A$ and $f_B$. **(d)** Evolution of the four oscillator frequencies when the frequency of external source A is swept. One after the other, the oscillators phase-lock to the external input when the frequency of the source approaches their natural frequency. In the locking range, the oscillator frequency is equal to the input frequency. **(e)** Experimental synchronization map as a function of the frequencies of the external signals $f_A$ and $f_B$. Each color corresponds to a different synchronization state. **(f)** Inputs applied to the system, represented in the ($f_A$, $f_B$) plane. Each color corresponds to a different spoken vowel and each data point corresponds to a different speaker.

The inputs to the neural network are encoded in the frequencies $f_A$ and $f_B$ of two fixed-amplitude microwave signals. Injected in a strip line fabricated above the active magnetic layers, they modify the dynamics of the oscillators through the radiofrequency magnetic fields they generate. Fig. 1d shows that when the frequency of one of the microwave sources is swept, each oscillator synchronizes to the source in turn. Indeed, when the frequency of the source gets close to the frequency of one of the oscillators, the strong signal of the source pulls the adaptable frequency of the oscillator towards its own. In the locking range, the frequency of the oscillator becomes equal to the frequency of the source[27]. The dark blue curve in Fig. 1c shows an example of spectrum measured when the two microwave inputs are injected simultaneously. Two peaks (in red) appear at frequencies $f_A$ and $f_B$ due to capacitive coupling with the strip line. In comparison to the spectrum without inputs (light blue curve), the emission peaks of oscillators 1 and 2 are pulled towards $f_A$, whereas oscillator 4 is phase-locked to input B (its emission peak merges with the one of input B at $f_B$). We label this synchronization configuration as (4B).

The possible outputs of the neural network, represented in different colors in Fig. 1e, are the different synchronization configurations that appear for different frequencies of the two input signals, keeping the dc currents through the oscillators fixed. Depending on the frequencies of inputs, zero (grey regions), one or two oscillators are phase-locked. For example, in the petrol blue region labelled (2A), oscillator 2 is synchronized to input A. In the white region labelled (1A, 3B), oscillators 1 and 3 are synchronized to inputs A and B respectively.

We now describe how this neural network can recognize patterns by classifying spoken vowels, which are naturally characterized by frequencies called formants[28]. We use as input data a subset of the Hillenbrand database[a] (provided as supplementary material) comprising seven vowels pronounced by 37 different female speakers, where each vowel is characterized by 12 different frequencies. Formant frequencies are typically comprised between 500 and 3500 Hz, therefore a transformation is needed to obtain input frequencies ($f_A$, $f_B$) in the range of operation of our oscillators between 325 and 380 MHz. As detailed in Methods, we obtain $f_A$ and $f_B$ through two different linear combinations of the 12 formant frequencies that fit the grid-like geometry of the oscillator synchronization maps. In the resulting map shown in Fig. 1f, each point corresponds to one speaker. The spread in frequency for each vowel indicates that each speaker has a different pronunciation. Our goal is to recognize the vowel presented as input to the oscillator network independently of the speaker. For this purpose, the scattered points corresponding to each vowel pronounced by different speakers should all be contained inside a different region of the oscillator synchronization map in Fig. 1e.

---

[a] *available at https://homepages.wmich.edu/~hillenbr/voweldata.html*

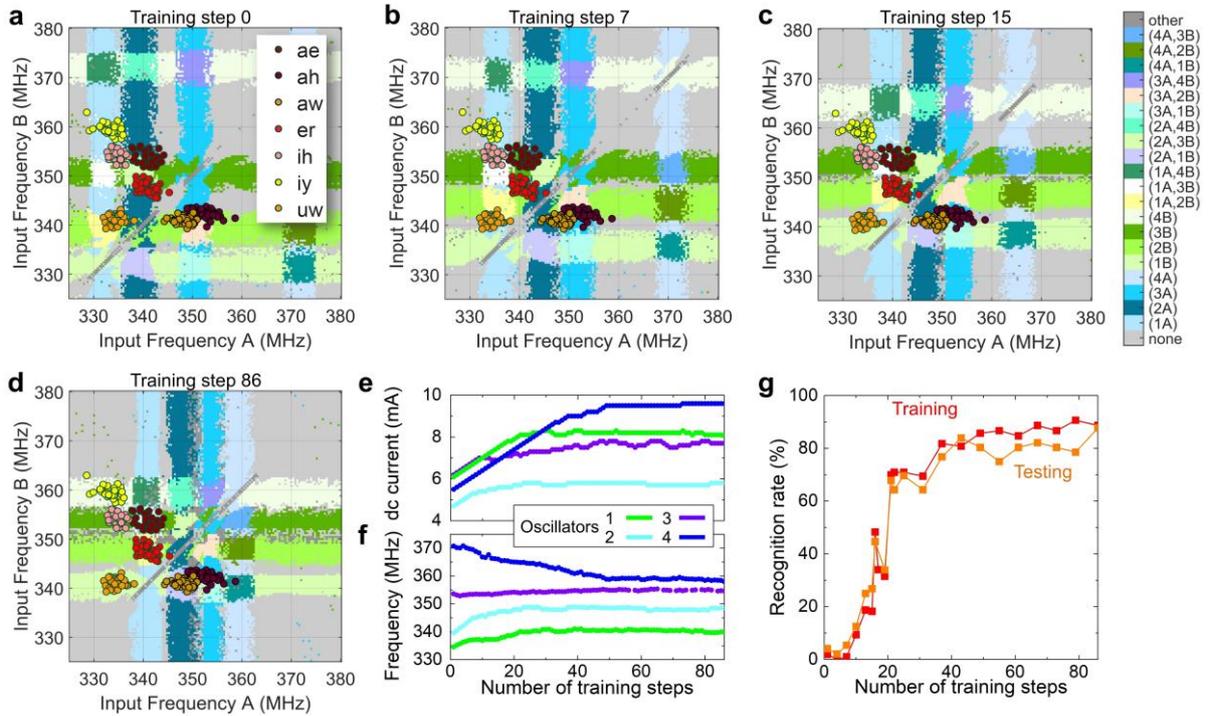

**Fig. 2. Learning to classify patterns by tuning the frequencies of oscillators.** (**a-d**) Experimental synchronization map as a function of the frequencies of the external signals, at different steps of the training procedure: (a) step 0 (b) step 7 (c) step 15 and (d) step 86. The colored dots represent the inputs applied to the oscillatory network: vowels pronounced by different speakers. Different vowels are in different colors. Video is provided as Supplementary Material. **(e)** dc current applied through each oscillator as a function of the number of training steps. **(f)** Frequency of each oscillator as a function of the number of training steps. **(g)** Recognition rates obtained with the set of data points used for training (red curve) and for testing (orange curve), as a function of the number of training steps.

As can be seen from Fig. 2a, in which the input vowel map and the oscillator synchronization map are superposed, initially, they do not coincide: the initial oscillator frequencies have been set randomly and are not adequate to solve the problem. The oscillatory neural network has to learn to perform the classification properly. During this training stage, the internal parameters of the network need to be finely tuned until each synchronization region encompasses the cloud of points corresponding to the vowel it has been assigned. For this purpose, we take advantage of the high frequency tunability of spin-torque nano-oscillators to modify the synchronization map by tuning the dc current through each oscillator, adapting a training algorithm first proposed in ref.[17]. We have developed an automatic real-time learning procedure involving a feedback loop between the experimental setup and the computer that controls it (see Methods). At each training step, we consecutively apply seven inputs ($f_A$, $f_B$) to the oscillators, one for each vowel, randomly picked between the different speakers. The oscillator emissions corresponding to each of the seven input microwave signals are recorded with a spectrum analyzer. A computer identifies the corresponding synchronization states (see Methods). If all the seven vowels have been correctly classified in their assigned synchronization regions of the map ($f_A$, $f_B$), the dc currents are not changed. If one or several vowels have not been correctly classified, dc currents in the oscillators are modified in order to bring the assigned synchronization regions closer to the corresponding input frequency pairs ($f_A$, $f_B$) and thus reduce the classification error (see Methods). In the next learning step, another set of seven vowels is applied and so on.

Fig. 2 shows synchronization maps obtained at different stages of the training process (Fig. 2a-d), together with the evolution of the dc currents applied to the oscillators (Fig. 2e), their frequencies (Fig. 2f) and the average recognition rates for the seven vowels (Fig. 2g) (see Supplementary Materials for a short video and html page[b] for an extensive video). After 48 training steps, an optimum is found, dc currents and frequencies stop evolving and the recognition rates stop increasing, signifying that the training process can be stopped. During training, we do not use all the vowels in the database. We always retain 20% of the vowels to test the ability of the system to recognize unknown data. The final recognition rates on the training and testing data sets reach values up to 89% and 88% respectively (Fig. 2g).

---

[b] Full video (3" 30'): https://youtu.be/IHYnh0oJgOA – Short video (20"): https://youtu.be/bbRqqcxc-po

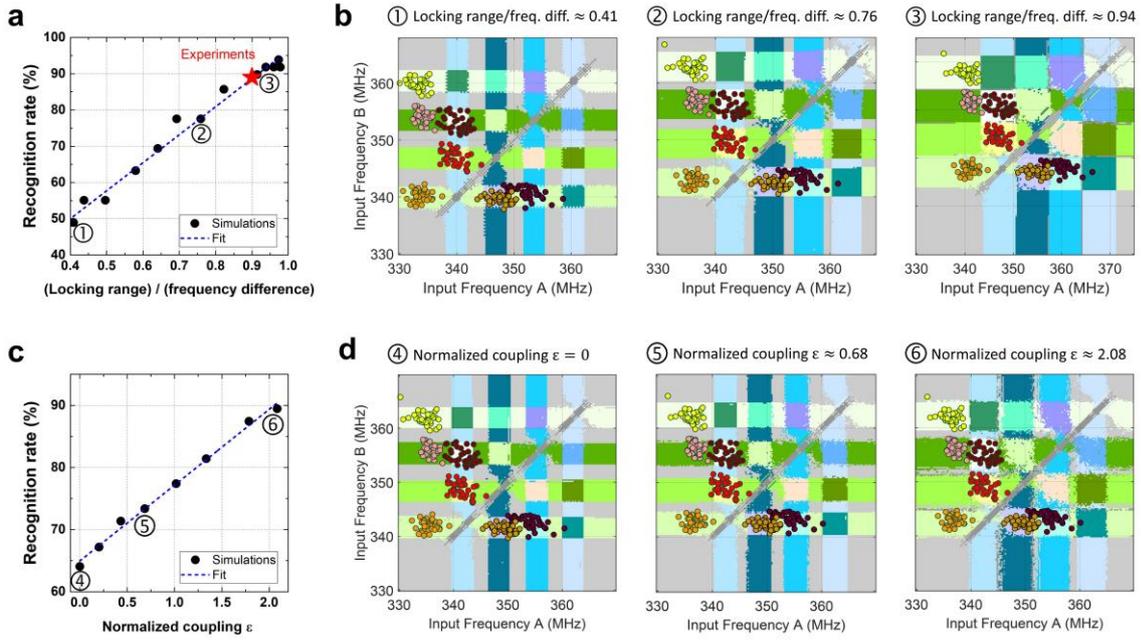

**Fig. 3. Comparing the recognition rates of experimental and ideal oscillators.** Simulations of vowel recognition with a network of four identical oscillators trained with the same procedure as in the experiments, in the absence of noise. The simulated oscillators differ only by a 2% mismatch in their natural frequencies. (**a**) Recognition rate on the training set (black circles) as a function of the average oscillator locking range normalized by the frequency difference between oscillators. The locking range is varied by modifying the oscillator frequency tunability. The blue dotted line is a linear fit to the simulation results. The red star indicates where experimental oscillators feature in this graph. (**b**) Synchronization maps simulated with the network of oscillators used in (a), for three different values of the normalized locking range. (**c**) Recognition rate on the training set (black circles) as a function of the mutual coupling between oscillators normalized by their coupling to the microwave inputs. The blue dotted line is a linear fit to the simulation results. (**d**) Synchronization maps simulated with the network of oscillators used in (c), for three different values of the normalized coupling.

We now interpret these experimental recognition rates by comparing them to the performances that can be achieved with ideal oscillators trained on the same task with the same learning process. For this purpose we model the oscillator dynamics with coupled van der Pol equations accounting for their collective magnetization coordinates (see Supplementary information)[20]. The simulated oscillators are noiseless and differ only by a 2% mismatch in their natural frequencies, analogous to the one observed experimentally. We first vary their ability to synchronize by modifying their frequency tunability (see Supplementary information). Black circles in Fig. 3a show the recognition rate of the ideal simulated network as a function of the average locking range of the oscillators normalized by their frequency difference. The recognition rate increases linearly with the oscillator locking ranges (see dotted blue linear fit in Fig. 3a). Indeed, as shown in the simulated maps of Fig. 3b, when the oscillator locking ranges increase, the regions of synchronization grow, thus encompassing and classifying an increasing number of points in each of the different vowel clouds. As shown in Fig. 3c and d, the mutual coupling between oscillators also enhances their locking ranges[27], leading to increased recognition rates when the mutual interactions increase. The red star in Fig. 3a pinpoints where the experimental result features in this graph. The experimental vowel recognition rate of 89% is close to the maximum recognition rate of 94% that can be achieved with the same neural network composed of ideal, noiseless oscillators. This high performance is due to the large experimental locking ranges resulting from the high tunability, coupling and low noise of the hardware spin-torque nano-oscillators.

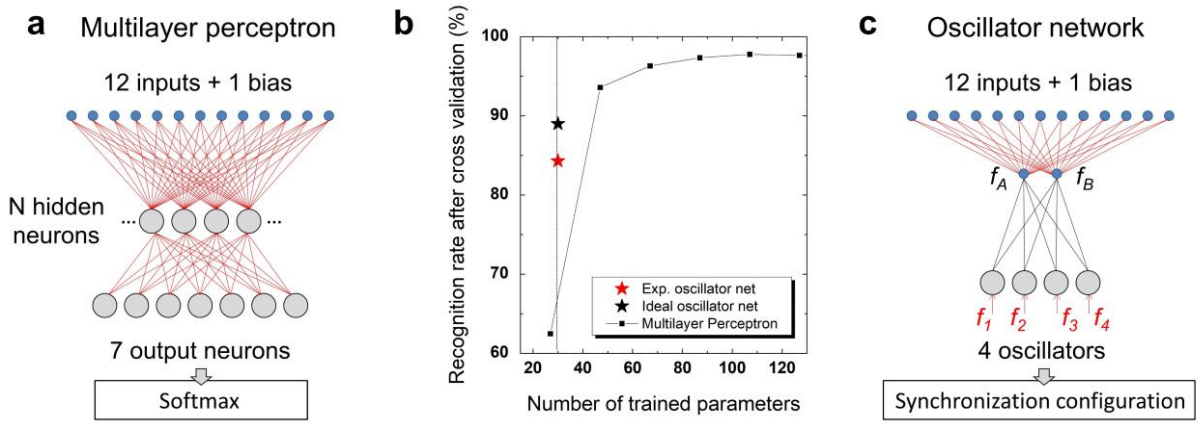

**Fig. 4. Benchmarking performances with classical neural networks. (a)** Flow chart of the simulated multilayer perceptron. The trained parameters are indicated in red. **(b)** Recognition rate obtained through cross-validation versus total number of trained parameters for the neural network in (A), in which the number of hidden neurons is varied. The red star corresponds to the experimental results with the network of spin-torque nano-oscillators. **(c)** Flow chart of the experimental oscillatory neural network. The trained parameters are indicated in red.

We then compare the dynamical oscillator-based neural network studied in this paper to more conventional forms of neural networks. For this purpose, we first extract a reference value for the experimental recognition rate by repeating the training procedure experimentally several times with different combinations of training and testing sets (see Methods). This cross-validation technique yields an average value of 84.3% for the experimental recognition rate on the testing set that we can compare to other neural networks performances. First, we consider a conventional, static, multi-layer neural network. This kind of network can achieve better-than-human recognition rates at complex tasks, such as image classification. This performance however, comes at the expense of the large number of parameters that need to be trained, a major hurdle for hardware implementation. Fig. 4b shows the recognition rate of a multilayer perceptron, trained in software through backpropagation on the same database as the experimental neural network, with 30,000 vowel presentations (see Methods). As illustrated in Fig. 4a, this network, composed of static neurons, takes as inputs the 12 formant frequencies characterizing each pronounced vowel. The hidden layer neurons receive a weighted sum of these inputs (plus a bias term). The output layer, with softmax activation functions, has seven neurons, one for each vowel class (see Methods). As can be seen in Fig. 4b, the recognition rate is excellent, reaching 97% when the number of trained parameters is large (synaptic weights illustrated in red in Fig. 4a). However, the performance rapidly degrades for small numbers of trained parameters, diving below 65% for 27 trained parameters. This result is quite general: as can be seen from Extended Data Fig. 2, state-of-the-art networks with feedback such as standard Recurrent Neural Network (RNNs) or Long Term Short Term Memory networks (LSTMs) have limited performance when the number of trained parameters is small. In contrast, the recognition rate of our experimental oscillatory neural network is over 84% for only 30 trained parameters: as illustrated in red in Fig. 4c, the 26 weights converting formants to inputs, and the currents through the oscillators. For an ideal, noiseless, oscillatory network, the success rate reaches 89% after cross validation. The network also learns rapidly (350 vowel presentations are used). This high performance with a small number of trained parameters comes from the combination of two phenomena: as shown in Fig. 3c the oscillatory network can do better than the sum of its individual components due to its complex, coupled, dynamical features and in addition, the oscillators collectively contribute to pattern recognition by synchronizing to the inputs. This result shows that the performance of hardware neural networks can be boosted by enhancing neuron functionalities beyond simple non-linear activation functions, through oscillations and synchronization.

In the future, such dynamical neural networks will have to be scaled up in order to solve challenging classification problems on software-benchmarked databases. Spin-torque nano-oscillators offer numerous advantages towards this goal. Their energy consumption is comparable or lower than CMOS oscillators, and contrary to the latter, their lateral dimensions can be scaled down to a few nanometers in diameter (a detailed comparison is presented in Extended Data Table 2). Their quality factor can exceed several thousands[26] and their natural frequency can be controlled by the aspect ratio of the magnetic dot from hundreds of MHz to several GHz in small pillars, opening the path to nano-oscillators assemblies with a wide range of natural frequencies[19]. In addition, their simple structure is similar to Spin-Torque Magnetic Random Access Memory cells, which means that they can be produced by billions on top of CMOS. Finally, their synchronization can be detected with CMOS circuits counting the number of oscillations[29], or measuring the additional dc voltages produced by the oscillators when they phase-lock (see Methods and Extended Data Fig. 3)[30]. Therefore the wide variety of possible magnetic and electric couplings offered by spintronics[21–24], the different ways of driving and controlling magnetization dynamics (spin-torques, spin-orbit torques, electric fields) can be exploited in the future to implement large scale hardware neural networks[15].

**Acknowledgements**

This work was supported by the European Research Council ERC under Grant bio*SPIN*spired 682955, the French National Research Agency (ANR) under Grant MEMOS ANR-14-CE26-0021 and by a public grant overseen by the ANR as part of the "Investissements d'Avenir" program (Labex NanoSaclay, reference: ANR-10-LABX-0035).

**Author contributions**

The study was designed by J.G. and D.Q., samples were optimized and fabricated by S.T. and K.Y., the main experiments were performed by M.R. and P.T., spin diode experiments were performed by P. T. and J. T., numerical simulations were realized by P.T., M. E., M.R., T. H. and D.V. All authors contributed to analyzing the results and writing the paper.

**Author information**

The authors declare no competing financial interests. Correspondence and requests for materials should be addressed to J.G. (julie.grollier@cnrs-thales.fr) and D.Q. (damien.querlioz@u-psud.fr).

# Methods

## A. Samples

Magnetic tunnel junctions (MTJs) films with a stacking structure of buffer/ PtMn(15)/ $Co_{71}Fe_{29}$(2.5)/ Ru(0.9)/ $Co_{60}Fe_{20}B_{20}$(1.6)/ $Co_{70}Fe_{30}$(0.8)/ MgO(1)/ $Fe_{80}B_{20}$(6)/ MgO(1) / Ta(8)/ Ru(7) (thicknesses in nm) were prepared by ultra-high vacuum (UHV) magnetron sputtering. After annealing at 360 °C for 1 h, the resistance-area product (RA) was $\approx$ 3.6 $\Omega\mu m^2$. Circular-shape MTJs with a diameter $\approx$ 375 nm were patterned using Ar ion etching and e-beam lithography. The resistance of the samples is close to 40 $\Omega$, and the magneto-resistance ratio is about 100 % at room temperature. The FeB layer presents a structure with a single magnetic vortex as the ground state for the dimensions used here. In a small region called the vortex core (of about 12 nm diameter at remanence for our materials), the magnetization spirals out of plane. Under dc current injection and the action of the spin transfer torques, the core of the vortex steadily gyrates around the center of the dot with a frequency in the range of 150 MHz to 450 MHz for the oscillators we used here.

## B. Database and inputs

In this study we classify seven spoken vowels with the oscillatory network. Spoken vowels are characterized by a set of frequencies called formants, that we obtain from a subset of the Hillenbrand database (*https://homepages.wmich.edu/~hillenbr/voweldata.html*) given in supplementary material. We use the first three formants ($F_1$, $F_2$ and $F_3$) sampled at four different times of the duration of the spoken vowel: at the "steady state" and at 20%, 50%, and 80% of the vowel duration respectively (*i.e.* 12 parameters in total). When one of these 12 parameters could not be measured or irresolvable formants mergers occurred, Hillenbrand *et al.* put a zero in this parameter in the database. For our study, we have removed the vowel utterances whose corresponding set of formants is not complete. Moreover, we use the same number of speakers for each vowel. The resulting formant database comprising 37 female speakers that we used is given in the Supplementary File "Formant-database.doc".

We perform two linear combinations of these formants in order to obtained two characteristic frequencies ($f_A$ and $f_B$) in the range of operation of the spin torque nano-oscillators (between 325 MHz and 380 MHz for the applied field value that we are using):

$f_A = A_1 \cdot F_1^{steady\_state} + B_1 \cdot F_2^{steady\_state} + C_1 \cdot F_3^{steady\_state} + D_1 \cdot F_1^{20\%} + E_1 \cdot F_2^{20\%} + G_1 \cdot F_3^{20\%} + H_1 \cdot F_1^{50\%} + I_1 \cdot F_2^{50\%} + J_1 \cdot F_3^{50\%} + K_1 \cdot F_1^{80\%} + L_1 \cdot F_2^{80\%} + M_1 \cdot F_3^{80\%} + N_1$

$f_B = A_2 \cdot F_1^{steady\_state} + B_2 \cdot F_2^{steady\_state} + C_2 \cdot F_3^{steady\_state} + D_2 \cdot F_1^{20\%} + E_2 \cdot F_2^{20\%} + G_2 \cdot F_3^{20\%} + H_2 \cdot F_1^{50\%} + I_2 \cdot F_2^{50\%} + J_2 \cdot F_3^{50\%} + K_2 \cdot F_1^{80\%} + L_2 \cdot F_2^{80\%} + M_2 \cdot F_3^{80\%} + N_2$

In order to choose the coefficients of the two linear combinations, we first record an experimental synchronization map that is used as a calibration of the network. The calibration map allows to assign a synchronization pattern to each vowel. Then, the linear transformation of the formants that best matches the data points of each vowel with its associated synchronization pattern is determined through fitting by least square regression. The coefficients used in the two linear combinations and the two frequencies $f_A$ and $f_B$ corresponding to each vowel are given in the Supplementary File "Formant-database.doc".

Once this calibration is done and the coefficients and characteristic frequencies are calculated, the dc currents are reset to random values to begin the learning experiment. Two fixed-amplitude microwave signals with frequencies $f_A$ and $f_B$ are used as inputs to the experimental network of coupled nano-oscillators.

C.  **Experimental set-up**

Extended Data Fig. 1 shows a schematic of the experimental set-up with the four coupled vortex nano-oscillators. A magnetic field of $\mu_0 H$ = 530 mT is applied perpendicularly to the oscillator layers to get an efficient spin transfer torque acting on the oscillator vortex core. A dc current is injected into each oscillator to induce vortex dynamics, which leads to periodic oscillations of the magnetoresistance, giving rise to an oscillating voltage at the same frequency than the vortex core dynamics. The four oscillators are electrically connected in series by millimeter-long wires. They are therefore coupled through the microwave currents they emit, and too far away to be coupled through the magnetic dipolar fields they radiate. Four dc currents ($I_{DC1}$, $I_{DC2}$, $I_{DC3}$, $I_{DC4}$) are supplied to the circuit by four different sources, allowing an independent control of the current flowing through each oscillator. The actual current flowing through each oscillator is given by $I_{STO1}=I_{DC1}$, $I_{STO2}=I_{DC2}+I_{DC1}$, $I_{STO3}=I_{DC3}+I_{DC2}+I_{DC1}$ and $I_{STO4}=I_{DC4}+I_{DC3}+I_{DC2}+I_{DC1}$ respectively; where $I_{STOi}$ corresponds to the current flowing through the $i^{th}$ oscillator. Two microwave sources are used to inject two external microwave signals with frequencies $f_A$ and $f_B$ and power $P$ = -9 dBm through a strip line, creating two microwave fields as inputs to the oscillator network. The amplitude of the generated magnetic field, set by Ampere's law, depends only on the cross section of the antenna (in addition to the distance between the strip line and the active magnetic layer of the oscillators). Therefore, the length of the antenna is only set by the number of oscillators it should cover. In our case, the strip line has a width of 2.5 µm and is fabricated 370 nm above the pillar (separated by an insulating layer). The resulting input microwave fields have an amplitude of 0.1 mT. They strongly affect the magnetization dynamics of the four oscillators, and thus the total microwave output emitted by the network. The microwave emissions are recorded with a spectrum analyzer. As can be seen in Fig. 1d, the input signals from the antenna can be detected in addition to the oscillator emissions due to capacitive coupling between the strip line antenna and the metallic electrodes connecting the oscillator. The analysis of the output, which depends on the frequencies of the microwave inputs, can therefore easily be used to classify the spoken vowels.

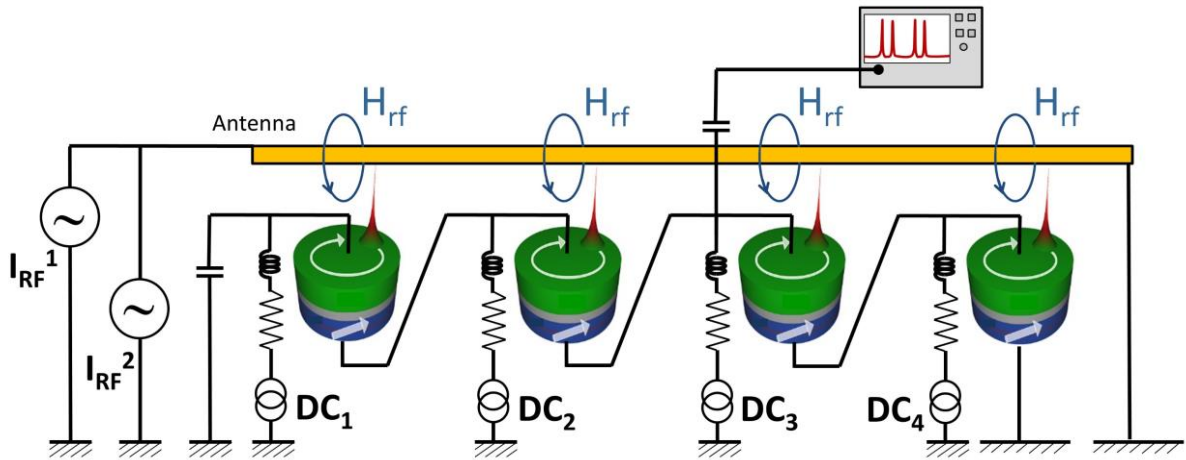

**Extended data Figure 1. Schematic of the experimental set-up.**

Each spectrum recorded with the spectrum analyzer is sent to the computer, where it is analyzed by a program in real time. The information we use as input to this program is: (i) the value of the two frequencies of the external microwave signals ($f_A$, $f_B$) and (ii) the oscillator frequencies at each dc current values in the absence of external microwave signals ($f_1^0$, $f_2^0$, $f_3^0$, $f_4^0$). The output data that we extract from each spectrum analysis are the four values of the oscillator frequencies in the presence of microwave inputs. Then, another program takes these oscillator frequencies to calculate the synchronization states and check if the applied vowel was properly recognized:

- If one of the detected frequencies coincides with the frequency of one of the external signals ($\pm 0.5$ MHz) we consider that the oscillator is synchronized to it.

- From this analysis, the synchronization pattern that corresponds to the input vowel is calculated.

- This is compared to the synchronization pattern initially assigned to that specific vowel to check if it was successfully classified or not.

If we are in the training procedure and the vowel is not properly classified, the on-line learning algorithm calculates how the four dc currents should be modified to reduce the recognition error, as described in section "Methods: Real-time learning algorithm". This information is then sent back to the experimental set-up, where the dc currents are automatically modified.

### D. Real-time learning algorithm

In this section, we present the supervised learning procedure that was applied to our spin-torque nano-oscillator network to learn to recognize different classes of input stimuli. Here these classes correspond to seven different spoken English vowels: "AE", "AH", "AW", "ER", "IH", "IY" and "UW". Initially, we assign a synchronization pattern to each class of vowel (column 2 in Extended Data Table 1).

| 7 different spoken vowel classes | Synchronization pattern | Associated frequency difference vector |
|---|---|---|
| "AE" | (1A, 3B) | $d_{ae} = \begin{pmatrix} f_A^i - \omega_1 \\ 0 \\ f_B^i - \omega_3 \\ 0 \end{pmatrix}$ |
| "AH" | (3A, 1B) | $d_{ah} = \begin{pmatrix} f_B^i - \omega_1 \\ 0 \\ f_A^i - \omega_3 \\ 0 \end{pmatrix}$ |
| "AW" | (2A, 1B) | $d_{aw} = \begin{pmatrix} f_B^i - \omega_1 \\ f_A^i - \omega_2 \\ 0 \\ 0 \end{pmatrix}$ |
| "ER" | (1A, 2B) | $d_{er} = \begin{pmatrix} f_A^i - \omega_1 \\ f_B^i - \omega_2 \\ 0 \\ 0 \end{pmatrix}$ |
| "IH" | (3B) | $d_{ih} = \begin{pmatrix} 0 \\ 0 \\ f_B^i - \omega_3 \\ 0 \end{pmatrix}$ |
| "IY" | (4B) | $d_{iy} = \begin{pmatrix} 0 \\ 0 \\ 0 \\ f_B^i - \omega_4 \end{pmatrix}$ |
| "UW" | (1B) | $d_{uw} = \begin{pmatrix} f_B^i - \omega_1 \\ 0 \\ 0 \\ 0 \end{pmatrix}$ |

**Extended data Table 1. Learning rule.** Spoken vowel class (column 1), synchronization pattern assigned to each vowel (column 2), and frequency difference vector between the spoken vowels and their associated patterns (column 3). The index i refers to the i[th] datapoint of a vowel class (i[th] speaker).

To have a perfect recognition of one class of vowel, all data points in the frequency input map that corresponds to this vowel (Fig. 1f) must be contained in their assigned synchronization pattern in the experimental map (Fig. 1e). If this is not the case, for each association spoken vowel-synchronization pattern we define a frequency difference vector with four components (one for each oscillator, see third column in Extended data Table 1) that will be used in the learning procedure.

Starting from a random map configuration (Fig. 1e), the automatic learning rule that we developed allows us to converge to a configuration where most data points for each vowel class are contained in their respective assigned synchronization pattern.

The learning rule works in the following way:

1) We present to the network a randomly chosen input data point $i$ belonging to one vowel class, by sending two microwave inputs with frequencies $f_A^i$ and $f_B^i$.

2) From the resulting spectra, we extract the frequencies of the four spin-torque oscillators $(\omega_1, \omega_2, \omega_3, \omega_4)$ in presence of the microwave inputs.

3) We determine the resulting synchronization configurations by comparing the oscillator frequencies to the input frequencies $f_A^i$ and $f_B^i$. Then, we compare the obtained synchronization configuration with the one assigned to this vowel.

4) For each vowel presented to the network, we define an associated frequency difference vector, which describes the frequency distance between the applied input and the assigned synchronization region. For instance, if the presented data point belongs to the vowel class "ae", we compute $\boldsymbol{d_{ae}} = \begin{pmatrix} (f_A^i - \omega_1) \\ 0 \\ (f_B^i - \omega_3) \\ 0 \end{pmatrix}$. If one of the two synchronization events assigned to "ae" has occurred, we only compute the frequency difference which corresponds to the other event. For instance, if oscillator 1 is correctly synchronized to external source $f_A^i$, then we compute only $\boldsymbol{d_{ae}} = \begin{pmatrix} 0 \\ 0 \\ (f_B^i - \omega_3) \\ 0 \end{pmatrix}$

5) We repeat steps 1) to 4) for all seven vowel classes.

6) We compute the sign of the vector sum of all seven associated frequency difference vectors $\boldsymbol{D}$:

$$\boldsymbol{D} = sgn(\boldsymbol{d_{ae}} + \boldsymbol{d_{ah}} + \boldsymbol{d_{aw}} + \boldsymbol{d_{er}} + \boldsymbol{d_{ih}} + \boldsymbol{d_{iy}} + \boldsymbol{d_{uw}}) = \begin{pmatrix} D_1 \\ D_2 \\ D_3 \\ D_4 \end{pmatrix}$$

7) Then, we compute the new dc current set $\begin{pmatrix} I_1' \\ I_2' \\ I_3' \\ I_4' \end{pmatrix}$, which will be applied to the four oscillators:

$$\begin{pmatrix} I_1' \\ I_2' \\ I_3' \\ I_4' \end{pmatrix} = \begin{pmatrix} I_1 \\ I_2 \\ I_3 \\ I_4 \end{pmatrix} + \mu \begin{pmatrix} D_1 \, sgn[\, (\frac{\partial \omega_1}{\partial I})_{I=I_1}] \\ D_2 \, sgn[\, (\frac{\partial \omega_2}{\partial I})_{I=I_2}] \\ D_3 \, sgn[\, (\frac{\partial \omega_3}{\partial I})_{I=I_3}] \\ D_4 \, sgn[\, (\frac{\partial \omega_4}{\partial I})_{I=I_4}] \end{pmatrix}$$

In this equation, $\mu = 0.1 \, mA$ is the learning rate of our algorithm. At each step, the applied dc current through each oscillator can be modified only by $\pm \mu$. Here $sgn[\, (\frac{\partial \omega_k}{\partial I})_{I=I_k}]$ represents the sign of the frequency evolution versus injected dc current of the $k^{th}$-oscillator at the value of current $I_k$. For this, the frequency – current dependence of each independent oscillator has been previously characterized.

Upon modifying the dc currents following this learning procedure, the oscillator frequencies change. This translates into a displacement of the synchronization patterns in the experimental synchronization map. (Fig. 2a-d)

8) We repeat all previous steps (step 1 to 7) $N$ times where $N$ is the total number of training steps. At each iteration, the synchronization map evolves towards an optimal configuration where the global frequency difference vector $\boldsymbol{d_{tot}} = \boldsymbol{d_{ae}} + \boldsymbol{d_{ah}} + \boldsymbol{d_{aw}} + \boldsymbol{d_{er}} + \boldsymbol{d_{ih}} + \boldsymbol{d_{iy}} + \boldsymbol{d_{uw}}$ is minimized. Upon increasing the number of training steps we observe an increase of the recognition rate until it saturates after step 48 reaching a value of 89 % (Fig. 2f). In our training experiment, we set the maximum number of training steps to $N = 87$, which corresponds to applying 3 times each of the 29 datapoints of the training database.

### E. Cross-validation procedure

Training was realized using 80% of the total number of vowels in the database. The testing procedure was done using the remaining 20% data points. The cross-validation technique allows estimating accurately the recognition performances of the network by repeating the training/testing procedure 5 times over distinct data point samples. Each time the selected data points used for testing are different: in the first (respectively second, third, fourth and fifth) cross-validation period, we use the first (respectively second, third, fourth and fifth) quintile (20%) of the data points for testing. The final recognition rate was obtained by averaging the testing recognition rates of the 5 cross-validation experiments. The same cross-validation procedure is used for all the neural networks (experimental and simulated).

### F. Comparison of spin-torque nano-oscillators to CMOS oscillators

Extended Data Table 2 compares features of CMOS and spin-torque nano-oscillators. "Vortex spin-torque oscillators" refer to the magnetic tunnel junctions used in this study. "10 nm spin-torque oscillator" refer to state of the art magnetic tunnel junctions currently used as memory cells.

| | Lateral dimensions | Energy / oscillation | Frequency | Power consumption | Ability to synchronize | References |
|---|---|---|---|---|---|---|
| CMOS neuron | > 30 µm | 265 pJ | 10 Hz | 2.65 nW | Yes | (31) |
| Scaled CMOS neuron | ≈ 7 µm | 50 pJ | 30 Hz | 1.5 nW | Yes | (32) |
| Accelerated CMOS neuron | ≈ 10 µm | 8.5 pJ | 1 MHz | 8.5 µW | Yes | (33) |
| CMOS ring oscillator | 6 µm | 6 fJ | 200 KHz | 1.2 nW | Unknown | (34) |
| CMOS ring oscillator | 6 µm | 33 fJ | 1.5 GHz | 50 µW | Unknown | (34) |
| CMOS ring oscillator | ≈ 300 µm | 1.4 pJ | 16 GHz | 23 mW | Yes | (35) |
| Vortex spin-torque oscillator | 300 nm | 3 pJ | 300 MHz | 1 mW | Yes | (36) |
| 10 nm spin-torque oscillator (projection) | 10 nm | 100 aJ | 10 GHz | 1 µW | Yes | (37) |

**Extended Data Table 2. Comparison of CMOS and spin-torque nano-oscillators for neuromorphic computing.**

### G. Comparison with a multilayer perceptron

In order to benchmark the results of the experimental oscillatory network, we first ran a standard multi-layer perceptron, schematized in Fig. 4a, on the same vowel database.

The network takes as inputs the 12 formants of a given vowel in a database and has seven outputs, one for each vowel class. We have varied the number of hidden neurons between 1 and 20 to evaluate the recognition rate as a function of the number of trained parameters. More precisely, each formant has been rescaled between -1 and 1 before being fed into the first layer of neurons. The neuron activation functions are tanh functions at the hidden layer, and softmax at the output layer: the outputs $z_i$ (i= 1 to 7) are defined as $z_i = e^{y_i}/\sum_{j=1}^{7} e^{y_j}$, where $y_j$ is the input to the output neuron $j$.

The output with the largest $z_i$ is taken as the vowel class corresponding to the input. We also tried ReLU activation functions, but they performed worse than tanh on this task.

For training the network we performed backpropagation, that is gradient descent over the negative log-likelihood (or cross entropy).

As in the experimental conditions, the samples are picked and presented randomly to the network. One learning iteration corresponds to one forward pass of a given sample through the network, its subsequent gradient evaluation and weight update. The learning rate has been tuned to obtain the best result. Weights and biases before learning were randomly sampled from a Gaussian of mean 0 and variance 0.01.

For each trial, we ran training over 100000 iterations to ensure convergence with a learning rate of 0.05. In practice, optimization techniques such as Root Mean Square Propagation or Adaptive Moment Estimation could be used to accelerate training. All results are reported in Fig.4b where we show the recognition rate after cross validation as a function of the number of trained parameters.

### H. Comparison with recurrent neural networks

In addition to the multilayer perceptron (MLP, Extended Data Fig. 2b), we also ran, on the same vowel database, a perceptron (Extended Data Fig. 2c), as well as a recurrent neural network (RNN, Extended Data Fig. 2d) and a Long-Short-Term-Memory recurrent neural network (LSTM, Extended Data Fig. 2e) with four hidden units.

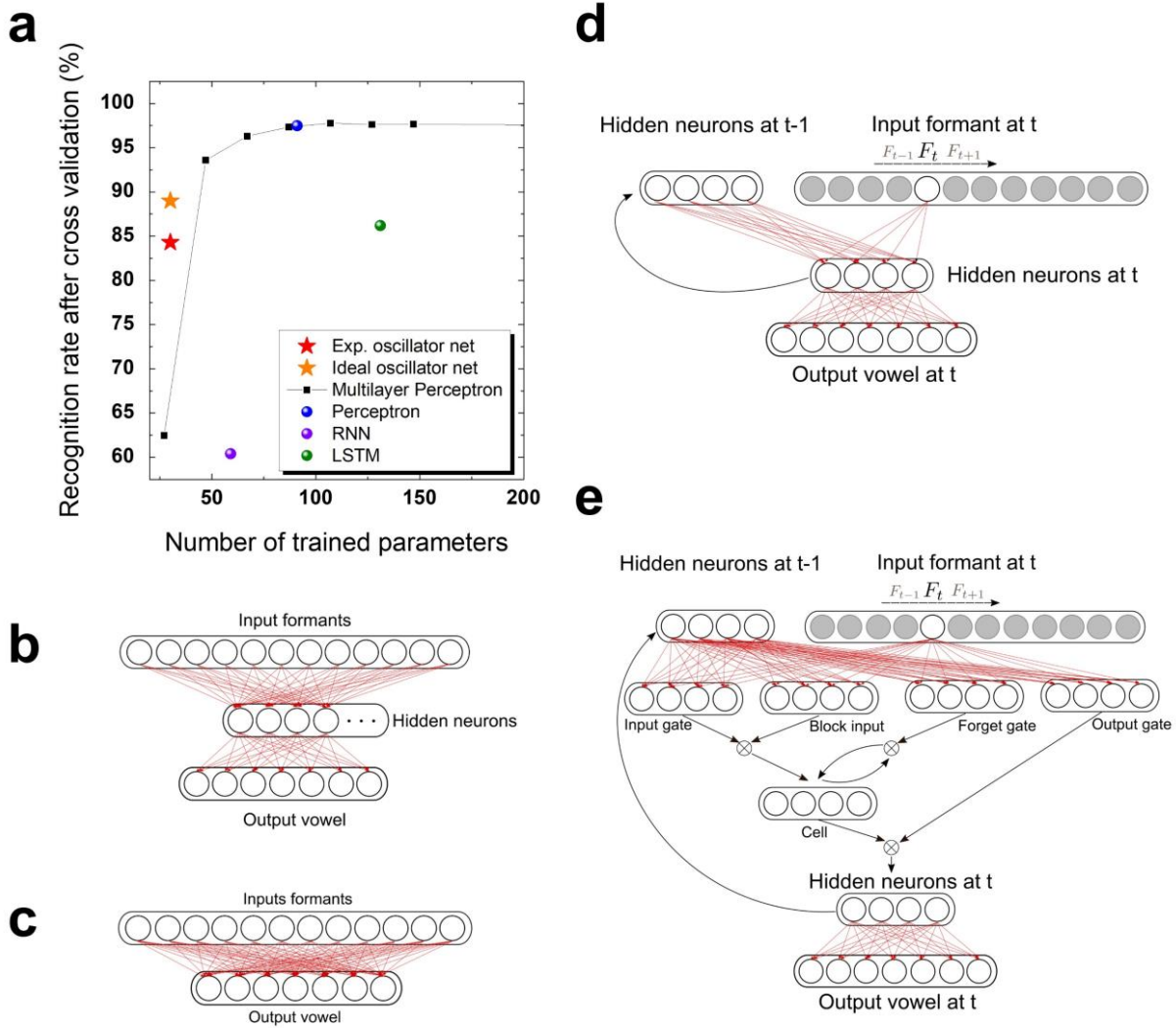

**Extended data Figure 2. Recognition rates obtained by different neural networks on the formant database. (a)** Recognition rates of different neural networks on the formant database as a function of the number of trained parameters. **(b-e)** Schematics of the simulated neural networks: **(b)** multi-layer perceptron, **(c)** perceptron, **(d)** recurrent neural network (RNN), **(e)** long-short term neural network (LSTM).

The procedure is similar to the multilayer perceptron. Formants are presented sequentially to the network which outputs a vowel once all of them have been swept through. Softmax activation functions were used at the output layer and tanh elsewhere. Outputs are encoded in a "one-hot" fashion: for example, the « ae » vowel (out of the 7 in total) is encoded by (1,0,0,0,0,0,0). We take the maximum activation value as the classification result. As in the experimental conditions, the samples are picked and presented randomly to the network. One learning iteration corresponds to one forward pass of a given sample through the network, its subsequent gradient evaluation and weight update. For each architecture, the choice of the learning rate has been tuned to obtain the best result. Weights and biases before learning were randomly sampled from a Gaussian of mean 0 and variance 0.01. No gradient inertia or learning rate adaptation technique was used. For the LSTM and the RNN, we ran training over 500000 and 1000000 iterations to ensure convergence with a learning rate of 0.01 and 0.0005 respectively. In practice, optimization techniques such as Root Mean Square Propagation or Adaptive Moment Estimation could be used to accelerate training. Due to the mini-batch size, gradient descent is highly stochastic and we average the test and training rates over the last 5000 iterations to obtain reliable training and error rate for a given trial. All results are reported in Extended Data Fig. 2a where we show the cross validation success as a function of the number of parameters learnt.

I. **Synchronization detection through oscillator rectified voltages**

In the present work, synchronization of the oscillators is detected using a spectrum analyzer, allowing a comprehensive understanding of the systems and of the physics of the oscillators.

In a final integrated system, simpler techniques could be used to detect synchronization of oscillators. A possibility is given in ref.[38]. Another method, involving less energy overhead, consists in exploiting the spin diode effect[39], which causes synchronized oscillators to generate a supplementary direct voltage[40]. Extended Data Fig. 3a-b illustrate this effect in one of our oscillators. The appearance of a rectified voltage measured between the oscillator electrodes (Extended Data Fig. 3a) coincides with the locking range (Extended Data Fig. 3b). The generated rectified voltage is proportional to the fraction of the external microwave current $I_{ext}$ flowing through the oscillator[40,41]. In our experiments $I_{ext}$ is small: the input microwave signals are sent though a strip line isolated from the oscillators, in a geometry minimizing by design the capacitive coupling between oscillator and strip line ($I_{ext}$ = 7.5e-3 $I_{stripline}$). As a result, the measured rectified voltages are small (~ 0.5 mV). In the future, these values can be increased up to several tens of mV by optimizing the coupling between oscillator and strip line. Indeed, as demonstrated experimentally, rectification effects due to oscillator phase locking can be large, with sensitivities reaching 75.4 mV for the generated dc voltage per µW of injected microwave power[41].

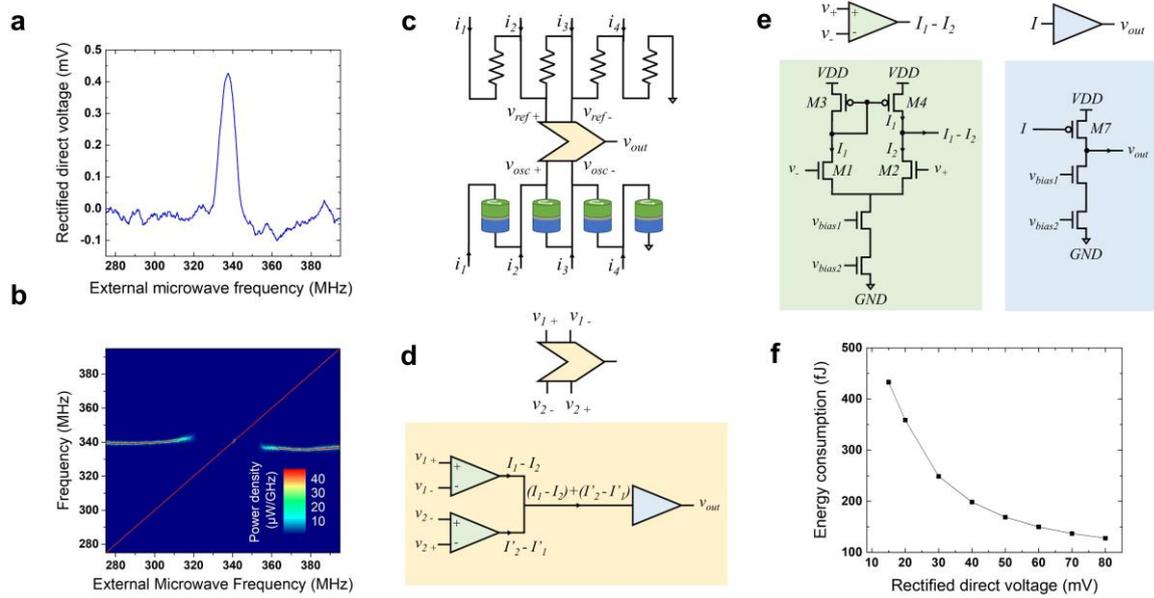

**Extended Data Figure 3**. **Synchronization detection by the spin-diode effect. (a)** Rectified direct voltage measured between oscillator electrodes when the external microwave signal is injected in the stripline above the oscillator and its frequency is swept. Here, the dc current through the oscillator is 5 mA, the magnetic field is 5850 Oe, and the injected microwave power is +1 dBm. **(b)** Oscillator spectrum emission measured during the same frequency sweep as (a). **(c)** Proposed differential measurement configuration for CMOS-based detection of synchronization-induced rectified voltages. **(d)** Two-stage CMOS circuit. **(e)** The first stage, composed of two differential amplifiers (green), is followed by a gain stage (blue). **(f)** Energy consumption of the CMOS circuit for one synchronization detection event, as a function of the amplitude of the generated rectified direct voltages.

We now present how synchronization detection through the resulting rectified voltages may be implemented in a final integrated circuit, using a differential method. We propose to use four reference resistors with the same resistance as the mean resistance of the nano-oscillators, and polarized in the same manner. Then, comparing the voltage across a nano-oscillator and the corresponding reference resistance allows detecting if the oscillator is experiencing synchronization (Extended Data Fig. 3c). We designed a simple two-stage Complementary Metal Oxide Semiconductor (CMOS) circuit to perform this comparison (Extended Data Fig. 3d-e). The first stage is composed of two differential amplifiers (voltage to current) in parallel. It is followed by a gain stage (current to voltage amplifier). The mismatch between the two amplifiers, a standard design technique, allows high gain. The output of the circuit is therefore a binary voltage, high if the oscillator is synchronized to the input signal, low otherwise. This voltage can be used directly by standard CMOS digital circuit to obtain the class of the input. In the circuit, bias voltages ($V_{bias1}$ and $V_{bias2}$) can be adjusted to vary the speed and power consumption of the circuit.

We simulated this circuit in transient operation using the Cadence Spectre SPICE simulator, a standard tool in commercial integrated circuit design, with the design kit of a 28 nanometer commercial CMOS technology, and optimized the bias voltages for minimal energy consumption, while retaining a response time of the circuit below 600 ns. Extended Data Fig. 3f shows the energy consumed by the detection circuit as a function of the rectified direct voltage due to synchronization, taking into account the whole transient of the detection. This energy can be low: it is below 200 fJ for rectified direct voltages above 50 mV, which can be achieved in structures optimized for spin diode effect[41]. For a full system, this detection has to be performed twice (we send two input signals), for the four oscillators, leading to a detection energy of 2 x 4 x 200fJ = 1.6pJ.

Using our current oscillators, this energy would be smaller than the energy dissipated by the oscillators and the reference resistors. By contrast, with scaled nano-oscillators (see Extended Data Table 2), this 1.6 pJ detection energy would become dominant.

It is interesting to compare this quantity with the energy consumption of a purely CMOS neural network, implementing the multilayer perceptron of Fig. 4a. Optimized CMOS neural networks compute in reduced precision, usually 8 bits integers, which allows low energy consumption[42]. Taking into account the arithmetic operations (sum and multiplications), in the same commercial 28 nanometer technology as the detection circuit that we implemented, we calculated that an 8-bits integer neural network implementing the second layer of the neural network of Fig. 4a consumes 2.2 pJ. We only took into account the second layer of the neural network, as it is the part implemented by the nano-oscillators. To obtain the energy estimation, we synthesized a Verilog description of a multiply and accumulate block, and computed its energy consumption with the Cadence encounter tools using appropriate value change dump files generated by the Cadence ncsim simulator.

These energy considerations show that on our tiny control system, a nano-oscillator based solution would provide an energy consumption slightly smaller than an optimized CMOS based solution. We expect that the full benefit of the oscillator system will appear in deep networks composed of many layers of spin-torque nano-oscillators. Indeed, cascading the synchronization states from one layer to the next can be achieved directly through oscillatory interlayer coupling and does not require synchronization detection. Only at the last layer will detection circuits be required to communicate their state to other circuits. Therefore, we expect that in a deep network of oscillators, the energy consumption will be largely dominated by the oscillator energy consumption which can be low for scaled oscillator as can be seen from Extended Data Table 2.

**Methods references**

**Data availability**

The datasets generated and analysed during this study are available from the corresponding authors on reasonable request.

## Supplementary information: numerical simulations

In this supplementary document, we present the numerical simulations that were performed to investigate the important features that oscillators should possess to classify accurately.

### 1. Model description

For simulating ideal oscillators, we consider the van der Pol model of non-linear dynamics that captures the essential features of spin-torque nano-oscillators coupled dynamics and can be generalized to other non-linear oscillators[1]. We consider four identical oscillators which only differ by a relative mismatch of 2% in their natural frequencies $f_0$, and which dynamics are modified by two microwave input signals. This leads to the following differential equations in polar coordinates $\begin{pmatrix} s_i \\ \theta_i \end{pmatrix}$, where index $i$ ($i$=1,2,3,4) represents the $i^{th}$ oscillator: (2)

$$\frac{ds_i}{dt} = -\alpha\omega_{0,i}\left(1 - \frac{I}{I_{th}} + Qs_i^2\right)s_i + F_e\cos\theta_i(\cos(\psi_{e,A} - \omega_{e,A}t) + \cos(\psi_{e,B} - \omega_{e,B}t)) + \varepsilon\, F_e\cos\theta_i \sum_{j=1}^{N=4} s_j\cos\theta_j$$

$$\frac{d\theta_i}{dt} = \omega_{0,i}(1 + N_0 s_i^2) + \frac{F_e}{s_i}\sin\theta_i(\cos(\psi_{e,A} - \omega_{e,A}t) + \cos(\psi_{e,B} - \omega_{e,B}t)) + \varepsilon\frac{F_e}{s_i}\sin\theta_i \sum_{j=1}^{N=4} s_j\cos\theta_j$$

In this equation, $\omega_{0,i}$ is the natural angular frequency of the oscillator, $\alpha = 0.013$ is the damping coefficient, $Q = 3.02$ is the nonlinear damping parameter, $I$ is the dc current injected in the oscillator, $I_{th} = 1\, mA$ is the threshold dc current of self-sustained oscillations of the magnetization, $N_0$ is the nonlinear frequency shift normalized by the natural angular frequency, $\omega_{e,A}$ and $\omega_{e,B}$ are the respective angular frequencies of the two external microwave inputs A and B, $\psi_{e,A}$ and $\psi_{e,B}$ are their relative phase shifts (Here $\psi_{e,A} = \psi_{e,A} = 0$), $F_e = 1.3 \times 10^{-3}$ is the coupling strength to each external microwave input signal A and B, and $\varepsilon$ the mutual coupling strength between oscillators, normalized by the coupling to the inputs.

## 2. Recognition performances

In this study we evaluate the impact of different oscillator characteristics (tunability and mutual coupling) on the classification performance of the network. With this purpose, some oscillator parameters are modified and a new optimized classification rate is calculated for each new set of material parameters. Each set of oscillator parameters corresponds to different oscillator behaviors and thus give rise to different synchronization maps. In particular the range of operation of the oscillators is modified and, in consequence, the linear combination previously applied to the formants to obtain two characteristic frequencies in the range of operation of the oscillators is no longer optimal. The linear combination of the formants should therefore be adapted for each point of Fig. 3a and c in order to determine the best recognition rates with the newly considered oscillator parameters. In this kind of network the recognition rate is optimized when:

(i) The free-running frequency difference between oscillators is similar:

$$|\omega_1 - \omega_2| \approx |\omega_2 - \omega_3| \approx |\omega_3 - \omega_4|$$

(ii) The width of the injection locking range of all 4 oscillators is similar:

$$\Delta_1 \approx \Delta_2 \approx \Delta_3 \approx \Delta_4$$

Thus, we first estimate which values of $\omega_i$ fulfil these requirements and we calculate the linear transformation of the formants whose final frequencies (inputs to the network) better fit the synchronization map expected from these $\omega_i$ and $\Delta_i$.

Finally, for each oscillator parameters and associated linear combination of the formants, we simulate numerically the learning process and find the optimum recognition rate.

Following this procedure, we study the influence of oscillator tunability and mutual coupling on the classification ability of our network.

In Fig. 3a-b, the oscillator locking ranges are modified by tuning their normalized non-linear frequency shift coefficient $N_0$ from 0.00 to 0.26 with a step of 0.02, in the absence of mutual coupling. The optimum recognition rate is calculated for each value of $N_0$. From the experimental frequency-current dependence the experimental nonlinear frequency shift can be obtained and averaged to $N_0^{exp} \approx 0.18$.

In Fig. 3c-d, the tunability is maintained fixed ($N_0$ = 0.08, corresponding to a value of locking range/frequency difference of 0.58 in Fig. 3a), and the normalized mutual coupling $\varepsilon$ is varied. The experimental value of the mutual coupling between oscillators coupling $\varepsilon$ is 1.6.

### 3. Synchronization maps

In the simulations of the synchronization maps, the dc currents applied to the 4 oscillators are kept constant and two external signals with a fixed external force are applied (we keep the same external force $F_e = 1.3 \times 10^{-3}$). We swept the frequency of two external sources A and B. Thus, each simulated synchronization map (see Fig.3) is constituted by 300x300=90 000 simulated points. These simulated points are independent from each other. This allows us to run simulations in parallel on GPUs.

Each simulated point in the map is calculated by numerically solving the system of coupled differential equations (1) using a fourth order Runge-Kutta scheme at $T = 0K$ (no thermal noise). Using the simulated cartesian trajectory and velocity of each vortex core in the dot plane $(x, y)$, the instantaneous frequency of each oscillator is extracted through the instantaneous angular evolution $\omega(t) = \frac{1}{2\pi}\frac{d\theta}{dt}$. The steady state frequency of each oscillator is obtained by computing the temporal average of the instantaneous frequency over only the last 20% of the simulated time trace. The synchronization between oscillators and microwave signals is detected by analyzing the frequency difference between oscillators and external sources:

If $|\omega_i - f_A| \leq f_{th}$, oscillator i and external source A are considered to be synchronized.

If $|\omega_i - f_A| > f_{th}$, oscillator i and external source A are considered to be not synchronized.

Where $f_{th}$ is a threshold set to $f_{th} = 0.5$ MHz.

**4 Evaluation of the injection locking range normalized by the frequency difference between oscillators**

The ratio between the injection locking range and the frequency difference used in Fig.3a is obtained by averaging the injection locking range of the 4 oscillators $\bar{\Delta} = \frac{1}{4}(\Delta_1 + \Delta_2 + \Delta_3 + \Delta_4)$ and the frequency difference between oscillators $\bar{\delta} = \frac{1}{3}(\delta_{12} + \delta_{23} + \delta_{34})$ where $\delta_{12} = |\omega_1 - \omega_2|$, $\delta_{23} = |\omega_2 - \omega_3|$, $\delta_{34} = |\omega_3 - \omega_4|$. The ratio defined in Fig. 3 A (denoted $\varrho$ here) is thus : $\varrho = \frac{\bar{\Delta}}{\bar{\delta}}$.

**Supplementary information references**